# An Improved Authentication & Key Exchange Protocol Based on ECDH for WSNs

Sina Baghbanijam
*Department of Electrical Enginieering*
*Iran University of Science and Technology*
Tehran, Iran
sina_baghbanijam@elec.iust.ac.ir

Hanie Sanaei
*Department of Electrical Enginieering*
*Tarbiat Modares University*
Tehran, Iran
hanie.sanaei@modares.ac.ir

Mahdi Farajzadeh
*Department of Electrical Enginieering*
*Tehran University*
Tehran, Iran
mahdi.farajzadeh@ut.ac.ir

*Abstract*— The widespread use of wireless sensor networks (WSNs) that are consisted of resource-constrained sensor nodes in communication with gateways in open-space environments and industries has highlighted the need for a secure yet fast communication protocol between users, gateways, and sensor nodes. Due to the properties of the network, elliptic-curve cryptography seems to be the most viable choice as it requires fewer resources than most other options. In this paper, we analyze the protocol suggested by Moghadam et al. which is based on ECDH (elliptic-curve Diffie-Hellman), and mention some of the flaws in their proposed authentication and key exchange protocol. Some attacks are also mentioned to further explain the shortcomings of their schema. Then a modified version of the protocol is proposed, analyzed, and checked against the same attacks as an informal security proof.

## I. Introduction

Ever since the introduction of IoT, the wireless sensor network has become even more discussed than it already was. Due to recent advancements in the field of electronics and telecommunications, it is now possible to have low-cost, low-power, yet high-performance sensors that are small enough to fit a wide range of applications. Wireless sensor networks, as the name suggests, are consisted of sensor nodes, gateway nodes, and interfaces. Interfaces provide users with a user interface to access the gateway node. Then the gateway node enables users to communicate with the sensor nodes while the sensor nodes gather environmental information and send it via an insecure channel to the gateway node. The wireless sensor network is deployed as a means to measure and monitor vibrations, temperature, pressure, and so on. This scenario is depicted in Fig. 1.

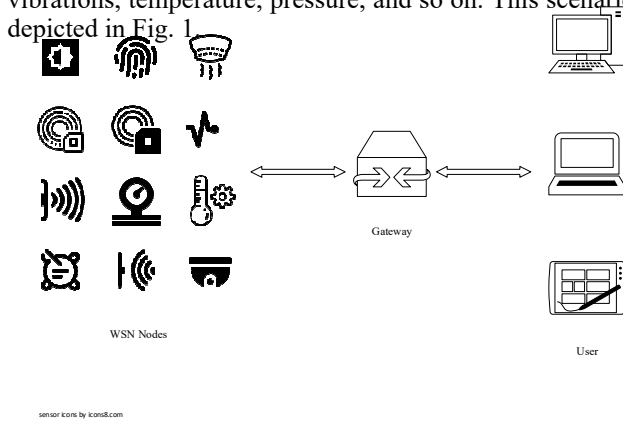

Fig. 1. The Discussed Scenario

Due to its widespread use in open-space environments [1], be it a small-scale business or a billion-dollar project, and the insecure channels used in-between the components, it is vital that a decent and secure protocol be used in communications. The said protocol must protect the transmissions against attacks to message integrity, authenticity, and confidentiality.

This paper is organized into six sections. In the next section, some of the basic concepts used in the paper are briefly mentioned. The Scheme proposed by Moghadam et al. [2] is analyzed in section III. In section IV the proposed authentication and key exchange protocol is disclosed and thoroughly explained. Next, the proposed protocol is analyzed in section V. Finally, a conclusion is drawn in section VI.

## II. Prerequisites

### A. Symmetric and Assymetric Cryptography

In the case of symmetric encryption, the same key which is considered a shared secret is used by the parties as a means of both encryption and decryption [3]. At times, it is assumed that the key is shared over a secure channel which is one of the main complications of using symmetric encryption and if not addressed properly, by each use of the key, some information might leak which could potentially accumulate and as a consequence put the message at risk. Even though key management issues are present in this kind of encryption, its considerably faster speed, lower overhead on the network, and less strain on the CPU make it a viable choice and is thus used along with asymmetric encryption more often than not.

On the other hand, asymmetric encryption utilizes a pair of mathematically related keys, a public and a private one [4]. The public key, available to everyone, is used to encrypt a plaintext message before sending it, but the private key is required to decrypt and read the encrypted message. Since the key pair is made by utilizing the concept of a trapdoor, the private key is not so easily derivable even by having access to the public key. Compared to symmetric encryption, the asymmetric one is considered more complicated and thus the needed time to complete the process is higher. Despite that, the fact that the private key is not shared and is kept secret offers some highly desired security advantages. It is also a considerably more scalable technique, unlike its symmetric counterpart.

Hence, asymmetric encryption is utilized first in the proposed protocol as a means of agreeing and sharing a session key between the parties, but then symmetric cryptography is called upon using the agreed session key for the duration of the session in which both parties make use of the agreed key for both encryption and decryption.

## B. Elliptic-Curve Cryptography

Elliptic-Curve Cryptography (ECC) is a kind of public-key cryptography based on elliptic curves over finite fields which consists of the points satisfying the equation:
$$y^2 = x^3 + ax + b$$
along with a distinguished point at infinity [5]. The coordinates in the above equation are chosen from a finite field of characteristic p not equal to 2 or 3 as they have a different equation each. It is also worthy to mention that it shall be non-singular, in other words, the condition $4a^3 + 27b^2 \neq 0$ shall check out. Two of the most used operations in elliptic curves are point sums and scalar multipliers. Based on the elliptic-curve discrete logarithm problem (elliptic-curve hard problem), if there were two points P and Q on the elliptic-curve which are selected in such a way that Q is a scalar multiplication of parameter K at point P ($Q = KP$), it would almost always be hard to obtain parameter K from points P and Q. The most promising feature of ECC is the lower key size required compared to most other public-key cryptography schemes like RSA, which in turn reduces storage and transmission requirements. This is because methods like index calculus do not apply to elliptic curves.

## C. Elliptic-Curve Diffie-Hellman Key Exchange Scheme

Elliptic-Curve Diffie-Hellman (ECDH) is a key agreement protocol that allows two parties who have a public and private key pair on an elliptic curve to share a secret over an insecure channel [6]. Considering that Alice is in hold of private and public key pair of (Sa, Xa) and Bob is in possession of private and public key pair of (Sb, Xb) which are based on an elliptic-curve of characteristic p, the two parties are able to agree on a session key of $Sa.Sb.p$ or some function of it since:
$$Sa.Xb = Sb.Xa = Sa.Sb.p.$$
It is noteworthy to mention that these key pairs can be randomly generated as needed and it is not necessary to have them prepared beforehand.

## III. ANALYSIS OF MOGHADAM ET AL.'S SCHEME

Moghadam et al. has proposed a mutual authentication and key exchange protocol to be used in WSNs. The proposed protocol is a modified version of the one Majid Alotaib [7] had proposed before it in which they suggested using a biometric fuzzy extractor to extract the biometric key of the user as a means of authentication. According to the authors, Majid Alotaib's proposed protocol is vulnerable to the stolen-verifier attack and contrary to the claims it lacks perfect forward secrecy. They have also dismissed the necessity of using biometric authentication means in the protocol due to its vulnerabilities.

To put it simply, their proposed protocol is divided into three phases. In the first one, aka the registration phase, the sensor and the user are registered. This process involves the utilization of an ID and a password along with a smart card for the user and a SID in the case of the sensor. Then there is the password change phase, in which the user is able to provide a new password since the protocol uses a password-based authentication and the use of a new password might be inevitable from time to time not to mention the need to change the password if it is compromised. Then there is the main body of the protocol, which is the authentication and key exchange phase. In this phase, the user inserts the smart card and provides the ID and password that are registered in the previous stages. After verification in the user's end, the request is sent to the gateway utilizing ECDH as a way to agree on a key to which all three parties contribute. While the protocol is rather compact and straight forward, there are some major issues that render it useless in a real application. Here, some of these flaws are stated:

A. The algorithm utilizes both random parameters and timestamps as a means of providing extra security esp. against replay attacks. Of course, utilizing timestamps is one of the easiest and simplest methods to prevent replay attacks, but only when it is used correctly. The problem is that the protocol sends the timestamp in its plain form instead of hiding or altering it by some means, which renders the whole concept redundant, and thus one can simply grab the message, alter the timestamp and send it at a later time.

B. The author considers SID a unique and private parameter, but there is a catch. Although SID is a private parameter and unique to a sensor that is utilized in the registration phase, it might be a shared parameter between multiple users that make use of the network. Not only that but also the fact that SIDs are mostly written somewhere on the device itself makes them rather insecure, and one has to avoid using it as much as he/she can, esp. as a method of authentication. That is because what gets authenticated is the knowledge of the SID and not the user requesting a communication link.

C. The author has stated that their proposed protocol is safe against DOS attacks but what is observed, opposite to this claim, is the fact that by sending a random message, the device performs multiple operations. One could easily utilize an altered version of the previously sent message by the user not to perform a replay attack but to simply disturb the service. Since the timestamp is implemented incorrectly, the message would pass the gateway and go to the sensor and back and back to the user again. If one had access to the SID as well, then it would be even worse since parameter A3 could be compromised as is explained below.

D. The parameter that the gateway authenticates upon user's request is A3 and based on the paper, $A3 = SID \oplus A_{2(x)}$ and as stated in their explanation, $A_{2(x)}$ is the selected base on ECDH. According to the NIST, each curve has a defined base as one of the predefined parameters and thus is known when using the elliptic curve. As stated earlier, Since the SID is used as a means to address the sensor with which the user is trying to converse, it might not be as private as one thinks it is. And that is the reason why parameter A3 which is being authenticated, not only fails to authenticate the user but is also prone to attacks since both of its making components are not particularly private.

E. It is stated in the paper that due to the risks and threats accompanied by biometric authentication, they have decided to omit the whole concept. While their concern is genuine [8], it might be too harsh to put the feature aside esp. when the majority of people prefer using their eyes or fingers instead of carrying a ton of cards around and be troubled by their loss from time to time. Not to mention that this feature could be useful in case a user forgot his/her password by accident, or it could even be used to further increase the security when facing stolen identity attacks by utilizing it along with the traditional password and smart card.

F. According to the authors, they have provided a password change phase while the protocol proposed by Majid Alotaib lacked this specific stage. In this phase, the parameters present in the card are updated but the parameters in the database remain unchanged. This could potentially lead to denial of service to a valid user.

G. Their protocol is also vulnerable to the privileged insider attack. As the gateway sends parameter D1 to the sensor, the malicious user could potentially intercept the message and since he/she knows parameter A2, then it would be possible to obtain KG since $D1 = KG \oplus A2$. KG is a private parameter saved by the sensor during the registration phase, and even a valid user should not be able to access it. Now, if someone obtained this parameter, then all the messages going from the gateway to the sensor would be compromised.

H. According to Kwon et al. [9], Moghadam et al.'s proposed protocol lacks perfect forward secrecy, and in case the gateway's private key gets compromised, the session key could potentially be revealed. This is in contrast to what they had claimed.

## IV. PROPOSED PROTOCOL

TABLE I. depicts the symbols used and their respective definitions in the proposed protocol. Next, there is the registration phase which is depicted in Fig. 2 and Fig. 3 for the user and the sensor respectively.

TABLE I. SYMBOLS AND DEFINITIONS

| Symbol | Definition |
| --- | --- |
| SID | Sensor Identity |
| ID | User Identity |
| PW | Password |
| BMP | Biometric Parameter |
| S | Gateway Private Key |
| X | Gateway Public Key |
| SK | Session Key |
| T | Time Stamp |
| H | Hash Function |
| $E_k$ | Encryption by Key k |

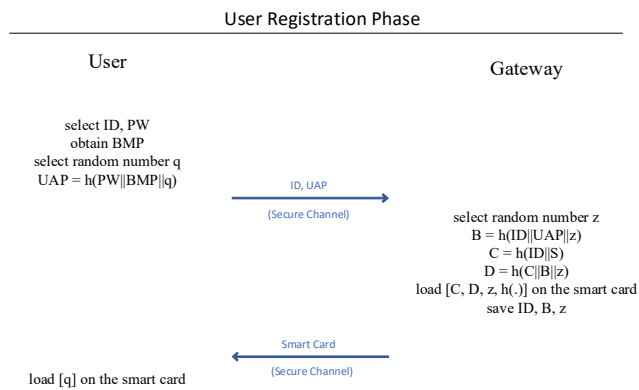

Fig. 2. User Registration Phase

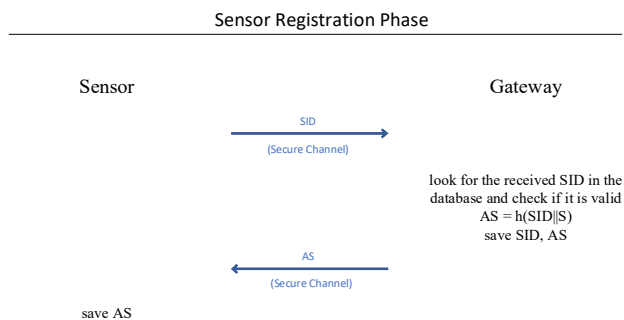

Fig. 3. Sensor Registration Phase

As for the user, he/she selects an ID and a password plus a biometric parameter that has already passed a hash function. The biometric parameter is assumed to be the hashed output of a device like a fingerprint or a retina scanner. PW and BMP together with a random number q are concatenated, and their hashed value is named user authentication parameter and sent along with the ID to the gateway via a secure channel. Gateway, in turn, selects a random number z and calculates parameters B, C, and D. B is the parameter that the gateway uses to authenticate the registered user later on and is formed by feeding concatenation of ID, UAP, and z to a hash function. C is the proof that the user has passed the gateway, and it can only be made by the gateway because of the secret parameter S and is made by hashing the concatenation of ID and gateway's private key. Parameter D, on the other hand, contains all the involved parameters and is used to authenticate a valid user after inserting the smartcard. It is made by concatenating C, B, and z and passing it through a hash function. Finally, the gateway loads parameters C, D, z, and the hash function to the smartcard and saves parameters ID, B, and z locally so that it is able to authenticate the user and provide perfect forward secrecy later on and then sends the smartcard back to the user who loads parameter q to the card after receiving it.

The process is simpler for the sensor. The sensor sends its SID to the gateway via a secure channel which searches the database for that specific sensor to see if the provided SID is valid or not. If it isn't, then it prompts for a valid SID and discards the one provided, but if it is indeed valid then it concatenates the provided SID with its private key S and passes it through the hash function, names it AS, and saves it along with SID so that it is able to authenticate the sensor later on and sends AS back to the sensor. Afterward, the parameter is saved locally by the sensor.

Since the proposed protocol is a password-based one, it is essential that one could change their passwords from time to time so as to keep an optimal security level. Therefore, a password change/update phase seems inevitable. Fig. 4 illustrates the proposed password change/update protocol.

As depicted, the ID, the PW and the BMP are first input by the user. After that, the parameter D* is made utilizing the input parameters and verified against the value of D present on the smartcard. If the two don't match, the operation halts, and another input is requested from the user. Otherwise, the user enters his/her new password and makes up a new UAP by concatenating the new PW, the BMP, and a new random number q. At that point a time stamp is obtained and parameters ID, B*, UAP, and the timestamp $T_1$ are encrypted utilizing $e_2$ as explained in section II part C. Next the PCR is sent along $e_1$ and $e_3$ to the gateway. The gateway first recognizes PCR and obtains a timestamp $T_2$ and proceeds to decrypting $e_3$ after calculating $e_2$ utilizing its private key S. Afterward, it checks the time stamp and parameter B* obtained by decrypting $e_3$ against the one saved in its database and drops the connection if any of the two don't match up.

Otherwise, it proceeds to make a new parameter B in the database and sends an encrypted version of it back to the user, confirming that it has indeed received the message and saved the new parameter so that the user can replace the old q and B with the new ones. The user does the aforementioned after authenticating the gateway by making sure the received message is indeed the correct parameter B. It is noteworthy to mention that if the gateway's message fails to deliver because of it being jammed or simply some channel error, the gateway does not replace the old parameter B for some time so that further authentication would still be possible and the valid user would not get locked out.

Then there is the actual authentication and key exchange protocol which is illustrated in detail in Fig 5. One of the best features of this protocol is that it has the minimum propagation delay possible as the request from the user passes through the gateway and heads to the sensor and comes back to the user through the gateway. The protocol is as follows: First the user inserts the smart card and inputs ID, PW, and the BMP and he gets verified on his side by computing parameters UAP, B and then D and checks if the made-up parameter based on the user's input is the same as the one on the smart card or not. If it weren't, then it prompts the user to try again, but if it checked out, then proceeds to the next step and utilizing random number a and timestamp $T_1$ it computes parameters $e_1$ = a.p and $e_2$ = a.X. p is the finite field characteristic as explained in section II. After this point, it encrypts parameters B*, SID, $T_1$ and sends it along $e_1$ to the gateway.

Upon receiving the message, the gateway selects timestamp $T_2$ and decrypts $e_3$ after remaking the encrypting key $e_2$ utilizing its private key and $e_1$. After decrypting and obtaining the timestamp in there, the gateway immediately checks it against the one made after receiving the message and if they don't match, then the message is discarded, but if they do, the received parameter B is checked against the one stored in its database as a means to identify the user as a valid one. If it is decided otherwise, the message gets discarded. Now, the gateway proceeds to relay the authenticated user's request to the sensor and to do so, it selects random numbers b and c and makes up AS = h(SID, s), $e_4$ = b $\oplus$ $e_2$ and $e_5$ = c.p. Next up parameters $SP_1$ = $e_4$ $\oplus$ AS and $SP_2$ = h($e_4$||SID||$T_2$) are made and sent along $T_2$ and $e_5$ to the sensor.

As the sensor receives the message, it selects timestamp $T_3$ and recovers $e_4$ since it has parameter AS saved in its memory in the registration phase. Then it makes up parameter $SP_2$ and checks it against the one sent by the gateway as a means of authenticating it because only the gateway has access to both the SID and parameter AS. If the two matches up, then the sensor checks message freshness and if either of the two fails, then the message gets discarded, but if not, the sensor considers it valid and proceeds to make the session key by first selecting a random number d. The session key is made by feeding the concatenation of the recovered $e_4$ with d.p to the hash function. Next the sensor makes up parameters $e_6$ = d.p and GP = h(SK||AS||$T_3$), and sends them along $T_3$ to the gateway.

The gateway first selects a time stamp $T_4$ and proceeds to build the session key utilizing $e_4$, random parameter c, and $e_6$. As mentioned in section II, both h($e_4$||c.$e_6$) and h($e_4$||d.$e_5$) are the same. Now the sensor builds parameter GP and checks it versus the one sent by the sensor to authenticate the message and proceeds to verify the message freshness. If neither of them fails, then parameters B, SKU which equals SK $\oplus$ z and $T_4$ are encrypted utilizing $e_2$ and sent over to the user.

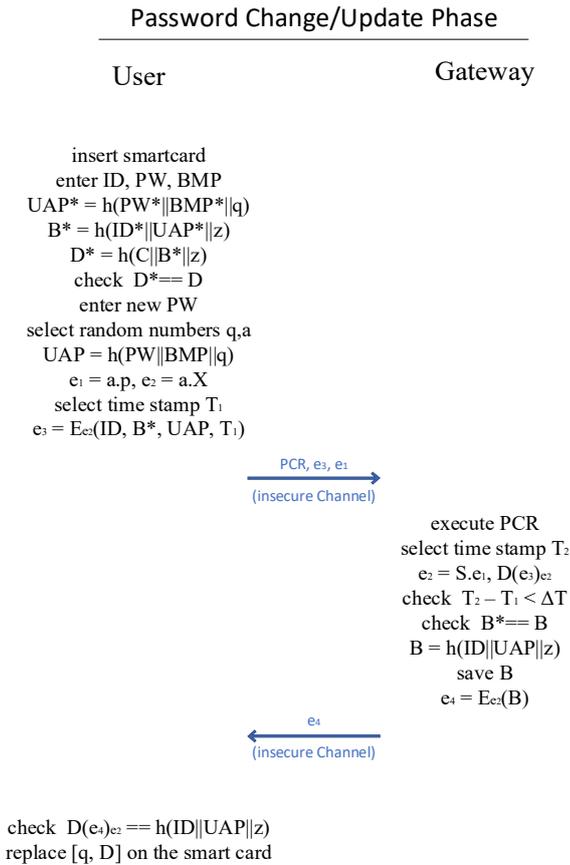

Fig. 4. Password Change/Update Phase

## Authentication and Key Exchange Phase

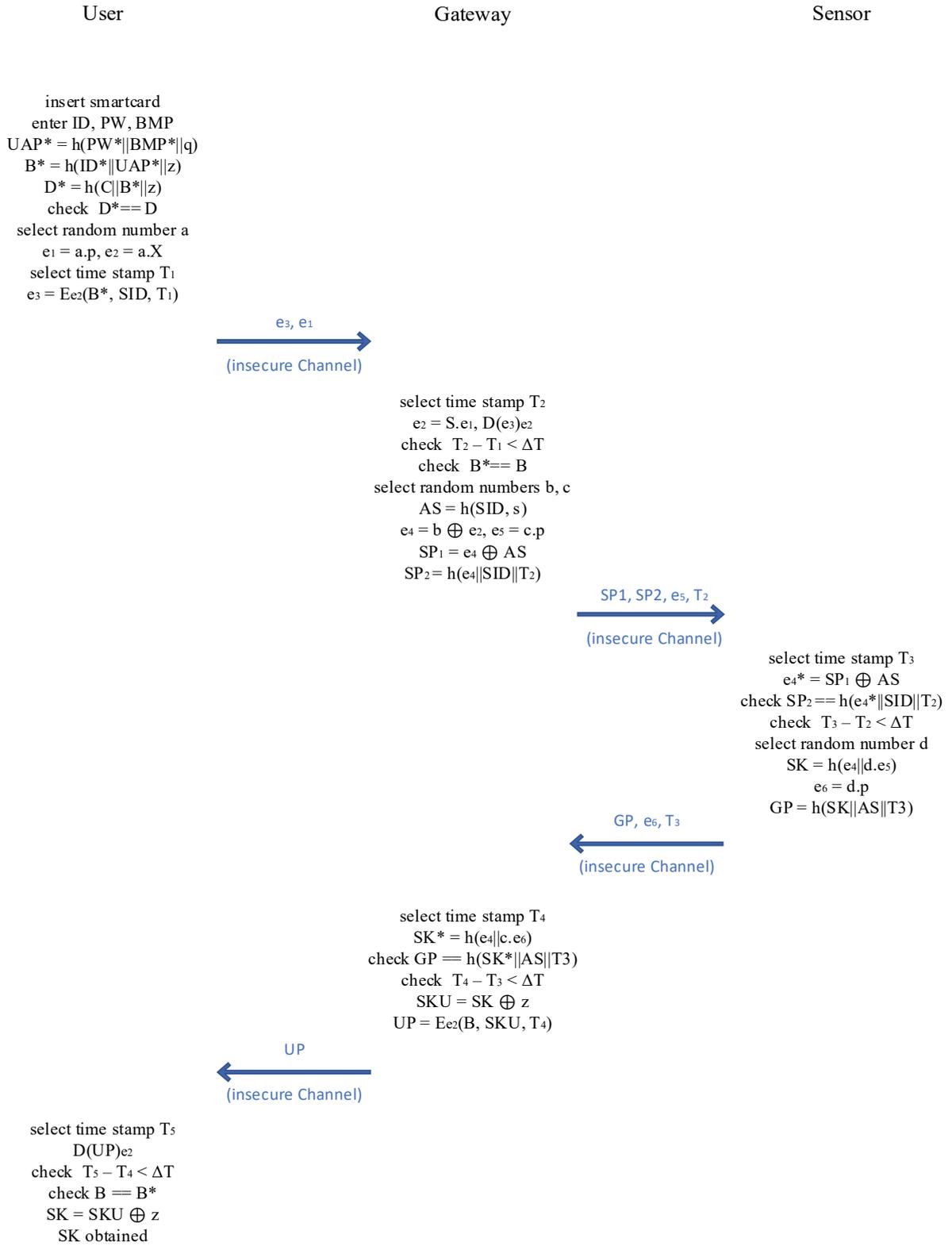

Fig. 5. Proposed Authentication and Key Exchange Protocol

The user selects a time stamp after receiving the message and proceeds to decrypt the message utilizing the very same key it sent the request to the gateway. After decryption, the user checks message freshness first, and later it weighs parameter B against the one sent by the gateway, and if the two were equal, then the user computes the session key and starts using it for the duration of the session.

## V. SECURITY AND PERFORMANCE ANALYSIS

As for the security analysis, an informal approach is preferred since utilizing a tool like Scyther [10] that is not updated in more than 5 years as a means of providing formal security proof is not recommended.

First, there was the issue with the timestamps which is now resolved by encrypting it between the user and the gateway, and checking SP2 and GP in the sensor node and the gateway respectively.

Next, in case a DOS attack occurs on the gateway or the sensor, it would either fail to pass either the time stamp or the authentication validity check, and the only processing power spent is for the calculation of e2 and decryption of e3 in the gateway and calculation of e4 in the sensor. The authentication parameter is also changed from A3 so that what gets authenticated is the user or the sensor itself. By doing so, one could potentially log any request in the gateway for further use as well.

The biometric parameter BMP is also implemented as an optional feature and could be utilized along with passwords as a means of improving the security of the protocol esp. when the password is compromised.

Perfect forward secrecy and resistance to privileged insider attack are provided by utilizing random parameter z when sharing SK with the user and utilizing random parameter b respectively.

A detailed password change/update protocol is also presented. Our approach does not require a secure channel as explained in section IV, which is a considerable upgrade compared to the one proposed by Moghadam et al.

TABLE II. depicts the overall security of the proposed protocol against some other similar protocols. The proposed protocol resists the attacks other similar protocols are vulnerable to. The user anonymity is removed due to its implementation being risky, and to grant the gateway the ability to authenticate each and every user and be able to log their every action which would be impossible were the users truly anonymous.

Next up to analyze the performance of the scheme, it would be logical to compare the computational costs of multiple similar protocols since the communication cost would not differ much when comparing protocols with the same number of handshakes. For the sake of comparison, the same 4-core 3.2 GHz CPU is hypothetically utilized [2], in which the execution time of the hash function $T_h$, ECC point multiplication $T_{ecc}$, and symmetric encryption/decryption $T_{sym}$ equal 0.00032s, 0.0171s, and 0.0056s respectively. The execution time of the XOR operation is considered negligible. TABLE III. shines some light on the performance of these schemes. What seems interesting is that the proposed protocol not only performs better but also offers extra security attributes when compared to that of Moghadam et al's.

TABLE II. SECURITY COMPARISON OF SIMILAR SCHEMES

| Protocols | [11] | [12] | [13] | [7] | [2] | Ours |
|---|---|---|---|---|---|---|
| Resist Replay Attack | Y | Y | Y | Y | N | Y |
| User Anonymity | N | Y | Y | Y | Y | N |
| Resist Stolen Verification Attack | N | Y | Y | N | Y | Y |
| Resist Stolen Smartcard Attack | N | Y | Y | N | Y | Y |
| Resist Impersonation Attack | N | N | N | Y | Y | Y |
| Resist Man in the Middle Attack | N | Y | Y | Y | Y | Y |
| Resist DOS Attack | Y | Y | Y | Y | N | Y |
| Perfect Forward Secrecy | Y | N | Y | N | N | Y |
| Resist Privileged Insider Attack | Y | Y | Y | Y | N | Y |

## VI. CONCLUSION

In this paper the protocol proposed by Moghadam et al. was analyzed and a multitude of issues were reported. Along with fixing the found issues, some considerable improvements were made to the overall security of the protocol and some new features were added to it. The biometric authentication is also implemented as an optional feature along other metrics. The password change/update phase is completed and explained in full detail, which can now be executed over insecure channels. Finally, the proposed scheme is compared to other similar protocols showing its merits and drawbacks. There are still some areas that could be the focus in later work. Among these are the better application of the biometric protocol along with earlier passwords to authenticate the user, and improvements regarding lowering the computational cost of authentication of the user by the gateway.

## ACKNOWLEDGMENT

We would like to thank Dr. Nasour Bagheri and Dr. Abolfazl Falahati for their guidance in analyzing the protocol.

TABLE III. PERFORMANCE COMPARISON OF SIMILAR SCHEMES

| Scheme | User | Gateway | Sensor | Total | Cost |
|---|---|---|---|---|---|
| [11] | $1T_h + 2T_{ecc}$ | $4T_h + 4T_{ecc}$ | $3T_h + 2T_{ecc}$ | $8T_h + 8T_{ecc}$ | 0.13936 |
| [13] | $6T_h + 3T_{ecc}$ | $6T_h + 1T_{ecc} + 1T_{sym}$ | $4T_h + 2T_{ecc} + 1T_{sym}$ | $16T_h + 6T_{ecc} + 2T_{sym}$ | 0.11892 |
| [2] | $5T_h + 4T_{ecc} + 2T_{sym}$ | $5T_h + 2T_{ecc} + 2T_{sym}$ | $3T_h + 1T_{ecc}$ | $13T_h + 7T_{ecc} + 4T_{sym}$ | 0.14626 |
| Ours | $3T_h + 2T_{ecc} + 2T_{sym}$ | $4T_h + 3T_{ecc} + 2T_{sym}$ | $3T_h + 2T_{ecc}$ | $10T_h + 7T_{ecc} + 4T_{sym}$ | 0.1453 |